\documentclass[12pt]{article}
\usepackage{epsf}
\begin{document}

\begin{titlepage}

\rightline{ISU-HET-01-3}
\rightline{hep-ph/0105006}
\rightline{April, 2001}

\begin{center}

{\Large\bf \mbox{\boldmath $K_L \rightarrow \pi^0 \gamma \gamma$} and the bound on
the CP-conserving \mbox{\boldmath $K_L \rightarrow \pi^0 e^+ e^-$}}

\medskip

\normalsize
{\large F. Gabbiani$^{}$ and G. Valencia$^{}$} \\
\vskip .3cm
$^{}$Department of Physics and Astronomy, Iowa State University,
Ames, IA 50011\\
\vskip .3cm

\end{center}

\begin{abstract}

It has been known for many years that there is a
CP-conserving component for the decay mode $K_L \rightarrow \pi^0
e^+ e^-$ and that its magnitude can be obtained from a measurement
of the amplitudes in the $K_L \rightarrow \pi^0 \gamma \gamma$ decay mode.
We point out that the usual description of the latter in terms of a single
parameter, $a^{\phantom{l}}_V$, is not sufficient to extract the
former in a model independent manner. We further show that there exist
known physics contributions to $K_L \rightarrow \pi^0 \gamma \gamma$
that cannot be described in terms of the single parameter
$a^{\phantom{l}}_V$. We conclude that a model independent analysis
requires the experimental extraction of three parameters.

\end{abstract}

\end{titlepage}

\section{Introduction}

The mode $K_L \rightarrow \pi^0 \gamma \gamma$ has been the subject of
intense study both as a test of chiral perturbation theory \cite{locpt}
and as the source of a CP-conserving amplitude for
$K_L \rightarrow \pi^0 e^+ e^-$
\cite{kpeeold}--\cite{reviews}. 
It has been known since the first experimental results appeared
\cite{earlyex} that lowest order ($p^4$) chiral perturbation theory is
not sufficient to explain simultaneously the observed rate and
spectrum. For some time now, it has become standard to use a
theoretical description which incorporates certain non-analytic terms
at next to leading order ($p^6$) \cite{unitam,unitco}, as well as one
parameter, $a^{\phantom{l}}_V$ \cite{unitco}.  This parameter arises
in vector meson dominance models for this decay \cite{avref}, but it
does {\it not} parameterize the most general analytic amplitude at
next to leading order in chiral perturbation theory, $p^6$. Instead,
at order $p^6$ the amplitude is described by three independent
parameters: $\alpha^{\phantom{l}}_1$, $\alpha^{\phantom{l}}_2$ and $\beta$ in the notation of
Cohen {\it et al.} \cite{unitco}.

Nevertheless, the parameterization of the amplitudes for $K_L \rightarrow
\pi^0 \gamma \gamma$ in terms of $a^{\phantom{l}}_V$ alone has
been retained in the literature. In this paper we wish to
point out that this is insufficient if one wants to extract a model
independent bound on the CP-conserving component of $K_L \rightarrow
\pi^0 e^+e^-$ from experiment. This is something which should be
considered by the forthcoming experimental analyses of the
mode $K_L \rightarrow \pi^0 \gamma \gamma$ by the KTeV
and NA48 collaborations.
Within the framework of chiral perturbation theory the new
data should be analyzed in terms of $\alpha^{\phantom{l}}_1$, $\alpha^{\phantom{l}}_2$ and $\beta$.
The issue of whether vector mesons dominate this decay mode is
an experimental question, and should {\it not} be an input in
the analysis of data. As a further motivation for the more general
fit, we show in this paper that there exists known physics,
the $f_2(1270)$, which affects the $K_L \rightarrow \pi^0 \gamma \gamma$
amplitude at a level comparable to that of vector mesons, and which
cannot be parameterized in terms of the single constant $a^{\phantom{l}}_V$.
It should be no surprise that the $f_2(1270)$ can play an important
role in this decay mode, given its prominence
in the reaction $\gamma \gamma \rightarrow \pi^0 \pi^0$ \cite{cball}.

Of particular importance is the determination of the
CP-conserving contribution to $K_L \rightarrow \pi^0 e^+ e^-$.
This contribution is completely dominated by one of the
two amplitudes present in $K_L \rightarrow \pi^0 \gamma \gamma$.
In fact, it is of phenomenological relevance only when it arises
from the amplitude in which the two photons are in a relative D-wave
\cite{sehgal}. For this reason an accurate determination of
both amplitudes is crucial. The model independent analysis
we advocate here permits the extraction of the necessary information
directly from the data, whereas
the usual analysis in terms of $a^{\phantom{l}}_V$ forces correlations
between the two $K_L \rightarrow \pi^0 \gamma \gamma$ amplitudes
which may or may not be present in the data.

\section{\mbox{\boldmath $K_L \rightarrow \pi^0 \gamma \gamma$} Amplitudes and Fit}

In this section we review the parameterization of the
$K_L \rightarrow \pi^0 \gamma \gamma$ amplitude with terms of
order up to $p^6$ in chiral perturbation theory and we compare
fits to the KTeV data from 1999 in terms of $a^{\phantom{l}}_V$ and in the
general parameterization.

The most general form of the $K\rightarrow\pi\gamma\gamma$ amplitude
contains four independent invariant amplitudes $A$, $B$, $C$ and $D$
and has been described in the literature before \cite{rafael}. For the case
of $K_L \rightarrow \pi^0 \gamma \gamma$, and in the limit of CP
conservation, only two of these amplitudes come into play:

\begin{eqnarray}
\lefteqn{{\cal M}(K_L(p^{\phantom{l}}_K) \rightarrow \pi^0(p^{\phantom{l}}_\pi)
\gamma(q_1)\gamma(q^{\phantom{l}}_2))\, =\,
{G_8 \alpha^{\phantom{l}}_{EM} \over 4 \pi}\epsilon_\mu(q^{\phantom{l}}_1)
\,\epsilon_\nu(q_2) \, \Bigg [
    A\,
    \left(q_2^\mu q_1^\nu - q^{\phantom{l}}_1\cdot
q_2 \, g^{\mu\nu}\right) } \nonumber \\
&&
+ 2 {B \over {m^2_K}}\,
\left(p^{\phantom{l}}_K\cdot q^{\phantom{l}}_1 \, q_2^\mu p_K^\nu
+ p^{\phantom{l}}_K\cdot q^{\phantom{l}}_2\, q_1^\nu p_K^\mu
- q^{\phantom{l}}_1\cdot q^{\phantom{l}}_2 \, p_K^\mu p_K^\nu -
p^{\phantom{l}}_K\cdot q^{\phantom{l}}_1\, p^{\phantom{l}}_K\cdot
q^{\phantom{l}}_2 \, g^{\mu\nu}\right) \Bigg ],
\label{gen}
\end{eqnarray}
where $G_8 = 9.1 \times 10^{-6}$~GeV$^{-2}$ and $\alpha^{\phantom{l}}_{EM} \approx
1/137$ is the usual electromagnetic fine structure constant.
In chiral perturbation theory with terms of order up to $p^6$,
the amplitudes $A$ and $B$ take the form \cite{unitco}:

\begin{eqnarray}
A(z) & = & 4 F\left(\frac{z}{r^2_{\pi}}\right) {a^{\phantom{l}}_1(z) \over z} +4 {F(z) \over z}
(1+r^2_{\pi}-z) \nonumber \\
& + &  {a_2 M^2_K \over {\Lambda^2_{\chi}}} \left\{ {4
r^2_{\pi} \over z} F\left(\frac{z}{r^2_{\pi}}\right) + {2 \over 3} \left(2 +
\frac{z}{r^2_{\pi}}\right) \left [{1 \over 6} + R\left(\frac{z}{r^2_{\pi}}\right)
\right] - {2 \over 3} \log {m^2_{\pi} \over M^2_{\rho}} \right.
\nonumber  \\
&-& 2 \frac{r^2_{\pi}}{z^2}(z+1-r^2_{\pi})^2 \left[ \frac{z}{12 r^2_{\pi}} +
F\left(\frac{z}{r^2_{\pi}}\right) + \frac{z}{r^2_{\pi}}
R\left(\frac{z}{r^2_{\pi}}\right)\right] \nonumber \\
&+& \left. 8 \frac{r^2_{\pi}}{z^2} y^2
\left[\frac{z}{12 r^2_{\pi}} + F\left(\frac{z}{r^2_{\pi}}\right) + \frac{z}{2
r^2_{\pi}} F\left(\frac{z}{r^2_{\pi}}\right) + 3 \frac{z}{r^2_{\pi}}
R\left(\frac{z}{r^2_{\pi}}\right) \right ] \right \} \nonumber  \\
& + &
\alpha^{\phantom{l}}_1 (z-r^2_{\pi})+\alpha^{\phantom{l}}_2, \nonumber \\
B(z) & = & {a_2 M^2_K \over {\Lambda^2_{\chi}}} \left\{ {4 r^2_{\pi} \over z}
F\left(\frac{z}{r^2_{\pi}}\right) + {2 \over 3} ( 10 - \frac{z}{r^2_{\pi}})
\left[ {1 \over 6} + R\left (\frac{z}{r^2_{\pi}}\right) \right] + {2 \over 3}
\log {m^2_{\pi} \over m^2_{\rho}} \right\} +
\beta,
\label{ampl}
\end{eqnarray}
where we use the kinematic variables
\begin{equation}
z  =  { \left( q_1 + q_2 \right)^2 \over M^2_K}\ ,\ \
y = {p_K \cdot(q_1-q_2)\over M^2_K},
\end{equation}
and $\Lambda_{\chi}$ $\approx$ $4\pi f_{\pi}$ $\approx$ 1.17 GeV.

This form for the two amplitudes does not correspond to
a complete calculation in
chiral perturbation theory at order $p^6$. It contains the complete
one-loop calculation of order $p^4$ \cite{locpt} and two types of terms
of order $p^6$. The first type consists of the non-analytic terms
in Eq.~\ref{ampl} that multiply the factors $a_2$ and $a_1(z)$.
The inclusion of these terms is inspired by dispersion
relations, and they originate in $p^4$ corrections to the $K \rightarrow
3 \pi$ amplitudes \cite{k3pi,dghkp}. The relevant constants which enter $a_1$ and
$a_2$ are extracted from an analysis of $K \rightarrow 3 \pi$ data.
The second type of term consists of the analytic terms that
arise from tree-level contributions from order $p^6$ chiral Lagrangians.
These contributions can be grouped into three unknown constants: $\alpha^{\phantom{l}}_1$,
$\alpha^{\phantom{l}}_2$ and $\beta$ corresponding to the three possible Lorentz
invariant forms which occur at order $p^6$ for the $K_L \pi^0 \gamma \gamma$
vertex \cite{unitco}. From the analysis of $K \rightarrow 3 \pi$ in
Ref. ~\cite{k3pi}, we have
\begin{eqnarray}
a^{\phantom{l}}_1(z) & = & 0.38 + 0.13 Y_0 - 0.0059 Y^2_0, \nonumber \\
Y_0 & = & {(z-r^2_{\pi} - {1 \over 3})\over r^2_{\pi}}, \nonumber \\
a^{\phantom{l}}_2 & = & 6.5,
\end{eqnarray}
with $r_\pi=m_\pi/M_K$. The loop form factors are given by \cite{unitco}
\begin{eqnarray}
F(z) & = & 1 - {4 \over z} \left[ \arcsin \left( {1 \over 2}\sqrt{z}
\right) \right]^2,
\, \qquad z \leq 4, \nonumber \\
& = & 1 + {1 \over z} \left( \log {1 - \sqrt{1 - 4/z} \over 1 + \sqrt{1 -
4/z}} + i \pi \right)^2, \, \qquad z \geq 4, \nonumber \\
R(z) & = & - {1 \over 6} + {2 \over z} \left[ 1 - \sqrt{4/z
- 1} \arcsin
\left( {1 \over 2} \sqrt{z} \right) \right], \, \qquad z \leq 4, \nonumber \\
& & - {1 \over 6} + {2 \over z} +  {\sqrt{1 - 4/z} \over z} \left( \log {1 -
\sqrt{1 - 4/z} \over 1 + \sqrt{1 - 4/z}} + i \pi \right), \,
\qquad z \geq 4. \nonumber
\end{eqnarray}

In the analysis of Ref. \cite{unitco}, which has become standard,
the three unknown constants were fixed in
terms of the contribution they receive from vector-meson exchange,
supplemented with a minimal subtraction ansatz:
\begin{eqnarray}
\alpha^{\phantom{l}}_1 &=& -4 a^{\phantom{l}}_V, \nonumber \\
\alpha^{\phantom{l}}_2 &=& 12 a^{\phantom{l}}_V -0.65, \nonumber \\
\beta &=& -8 a^{\phantom{l}}_V - 0.13,
\label{cohenan}
\end{eqnarray}
and this form has been used, for example, by KTeV \cite{ktev} to fit
their data with $a^{\phantom{l}}_V = -0.72\pm 0.05 \pm 0.06$.
In Eq.~\ref{cohenan} $\beta$ is no longer independent from
$\alpha_{1,2}$; therefore it is clear that this ansatz
introduces model-dependent correlations between the $B$ amplitude
(the one responsible for a large CP-conserving $K_L \rightarrow
\pi^0 e^-e^-$), and the $A$ amplitude which dominates the $K_L \rightarrow
\pi^0 \gamma \gamma$ mode, but which does not contribute significantly to
$K_L \rightarrow \pi^0 e^+ e^-$.

\begin{figure}[!htb]
\begin{center}
\epsfxsize=15cm
\centerline{\epsffile{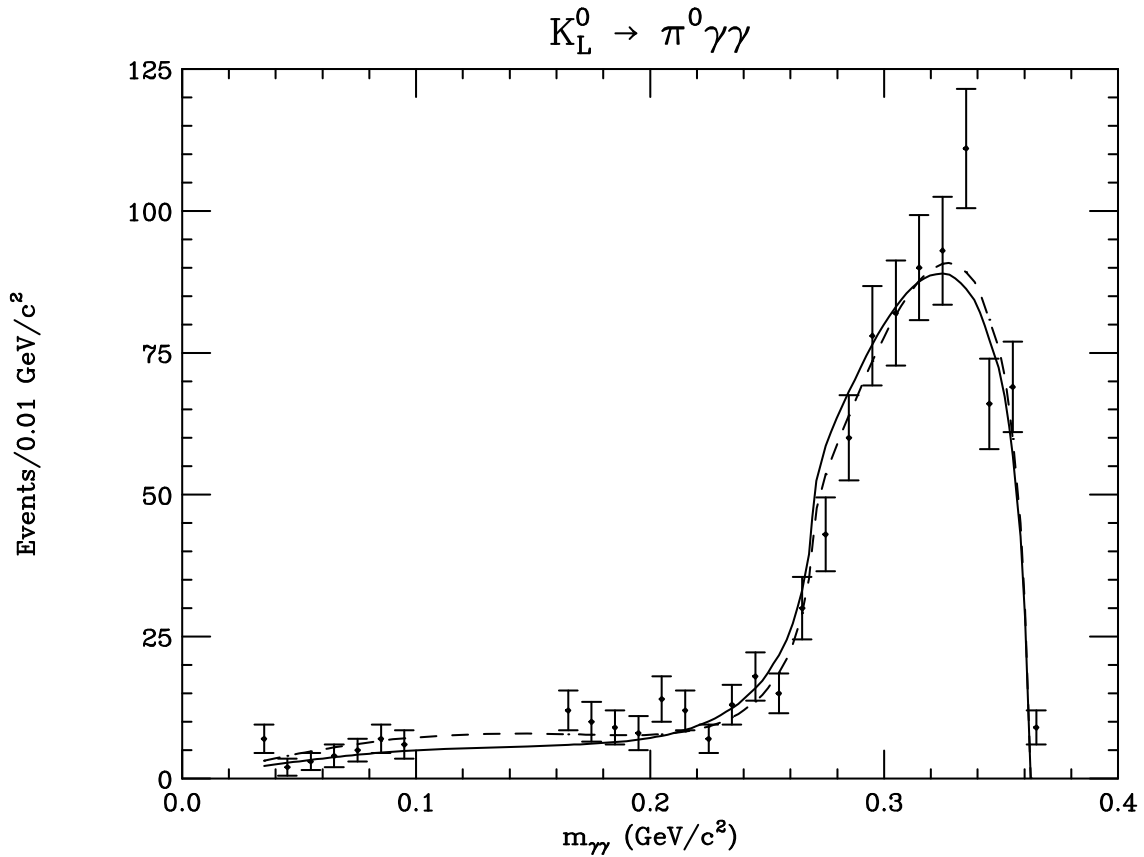}}
\end{center}
\caption{Two different fits to the data from Ref. \cite{ktev}, as
explained in the text. The solid line is a one-parameter fit
corresponding to Eq. \ref{avfit}, the dashed line is the three-parameter
fit shown in Eq. \ref{3pfit}.}
\label{fig:fit}
\end{figure}

In Fig.~\ref{fig:fit} we reproduce the data from Ref. \cite{ktev} as
can be read from their published paper. We superimpose on the data the
best fit we obtain in terms of the parameter $a^{\phantom{l}}_V$ as a solid
line. Our fit gives $a^{\phantom{l}}_V = -0.95$ with a $\chi^2/dof = 46/27$, which
corresponds to
\begin{eqnarray}
\alpha^{\phantom{l}}_1 &=& 3.8, \nonumber \\
\alpha^{\phantom{l}}_2 &=& -12.0, \nonumber \\
\beta &=& 7.5.
\label{avfit}
\end{eqnarray}

Notice that our value for $a^{\phantom{l}}_V$
is not the same value quoted by Ref.~\cite{ktev} because
we do not have access to the raw data and hence we have not taken into
consideration any background or detector issues. Nevertheless, we
feel that it is fair to compare this fit to our best three-parameter
fit obtained in the same way. This one is presented in Fig.~\ref{fig:fit} as
the dashed line, and corresponds to
\begin{eqnarray}
\alpha^{\phantom{l}}_1 &=& 0, \nonumber \\
\alpha^{\phantom{l}}_2 &=& 1.7, \nonumber \\
\beta &=& -5.
\label{3pfit}
\end{eqnarray}
For this fit we obtain a $\chi^2/dof = 37/25$, slightly better than
Eq.~\ref{avfit}. Clearly it is up to the experimentalists to present a
complete best fit to the data using the general form, Eqs.~\ref{gen},
\ref{ampl}, and taking into consideration all the experimental
issues\footnote{We proceed keeping the branching ratio fixed to the one
measured by KTeV \cite{ktev}, $(1.68 \pm 0.07 \pm 0.08) \times
10^{-6}$, in the normalization of our fits. Eventually the parameters
extracted from all our fits yield branching ratios very close to the
experimental one and well within its errors.}. However, it should be
clear from Fig.~\ref{fig:fit} that even though the current data is
consistent with the vector dominance assumption, it cannot rule out
other scenarios. In fact, our three parameter best fit is not
consistent with the vector meson dominance assumption. In the Appendix
we explore the significance of our fit showing the range allowed
for its three parameters within one sigma from our best $\chi^2$. This
provides an estimate of the errors involved.

\begin{figure}[!htb]
\begin{center}
\epsfxsize=6cm
\centerline{\epsffile{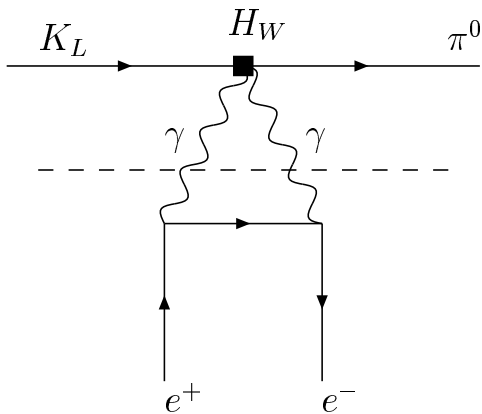}}
\end{center}
\caption{Contribution from the on-shell two-photon intermediate state to $B_{CP}(K_L
\rightarrow \pi^0 e^+ e^-$).}
\label{fig:cutf}
\end{figure}

Although the two types of fit are indistinguishable as far as
describing the $K_L \rightarrow \pi^0 \gamma \gamma$ spectrum, they
result in completely different predictions for the unitarity bound on
the CP-conserving contribution to $K_L \rightarrow \pi^0 e^+ e^-$. To
evaluate it we need to calculate the absorptive contribution from the
on-shell two-photon intermediate state to $K_L \rightarrow \pi^0 e^+
e^-$, as depicted in Fig.~\ref{fig:cutf}.

\noindent This yields the following bounds on the CP-conserving
part of $B_{CP}(K_L \rightarrow \pi^0 e^+ e^-)$:

\begin{equation}
B_{CP}(K_L \rightarrow \pi^0 e^+ e^-) \geq \cases{
2.3 \times 10^{-12} & {\rm VMD} \cr
3.4 \times 10^{-12} & {\rm three-parameter~fit} \cr
}
\label{comp}
\end{equation}

The above contribution is not the full absorptive part since
there is a further cut due to on-shell pions. Moreover, the full
CP-conserving amplitude includes a contribution from the dispersive
part of the amplitude, with off-shell photons (and pions).
The general form of the amplitude is
\begin{equation}
\label{K}
{\cal M}_{CP} (K_L \rightarrow \pi^0 e^+e^-) =
G_8 \alpha^2_{EM} K p^{\phantom{l}}_K \cdot (k_{e^+} - k_{e^-}) (p^{\phantom{l}}_K +
p^{\phantom{l}}_{\pi})^{\mu}
\overline{u} \gamma_{\mu} v,
\end{equation}
where $K$ is the result of the loop calculation and the extra
antisymmetry under $k_{e^+} \leftrightarrow k_{e^-}$ is a reflection
of the properties under a CP transformation. Introducing a form factor
to regularize the virtual photon couplings, an expression
for $K$ \cite{doga} is obtained:

\begin{equation}
K = {B(x) \over {16 \pi^2 m^2_K}} \left [
{2 \over 3} \log \left({m^2_{\rho}} \over {-s}\right)
- {1 \over 4} \log \left({-s} \over {m^2_e}\right)
+ {7 \over {18}}
\right ],
\end{equation}

\noindent where $s = \left( k_{e^+} + k_{e^-}\right)^2$.
The log factor is of course expected, since the photon absorptive
part comes from the expansion $\log (-s) = \log s + i\pi$.
This representation of the amplitude leads to CP-conserving branching
ratios:

\begin{equation}
B_{CP}(K_L \rightarrow \pi^0 e^+ e^-) = \cases{
4.8 \times 10^{-12} & {\rm VMD} \cr
7.3 \times 10^{-12} & {\rm three-parameter~fit} \cr
}
\label{comp2}
\end{equation}

\section{Resonance Models for \mbox{\boldmath $\alpha_1,\, \alpha_2$ and $\beta$}}

In this section we present the contributions of scalar and tensor
mesons to the parameters $\alpha^{\phantom{l}}_1$,
$\alpha^{\phantom{l}}_2$ and $\beta$. We will be able to show that
the tensor meson $f_2(1270)$, in particular, can contribute at a level
comparable to that of vector mesons and yet produce a different
pattern for the three constants.

We have chosen to follow the notation of \cite{unitco}, where the
parameters $\alpha^{\phantom{l}}_1$, $\alpha^{\phantom{l}}_2$ and
$\beta$ are defined by the expression for the invariant amplitudes
$A$ and $B$ as in Eq.~\ref{ampl}. It is convenient to relate these
parameters to the three Lorentz invariant couplings that can be
derived from a chiral Lagrangian at order $p^6$. Writing these
couplings as
\begin{equation}
{\cal L} = {G_8 \alpha^{\phantom{l}}_{EM} \over 4 \pi} \biggl(
c_1 K_L \pi^0 F^{\mu\nu}F_{\mu\nu} +
{c_2 \over M_K^2} \partial^\alpha K_L \partial_\alpha \pi^0
F^{\mu\nu}F_{\mu\nu} +
{c_3 \over M^2_K} \partial_\alpha K_L \partial^\beta \pi^0
F^{\alpha\mu}F_{\mu\beta} \biggr),
\label{count}
\end{equation}
where $F_{\mu\nu}$ is the usual electromagnetic field strength
tensor, one finds that
\begin{eqnarray}
\alpha^{\phantom{l}}_1 &=& -2 c_2 + {c_3 \over 2}, \nonumber \\
\alpha^{\phantom{l}}_2 &=& 4 c_1 + 2 c_2 +{c_3 \over 2}, \nonumber \\
\beta &=& - c_3.
\label{cpara}
\end{eqnarray}

The couplings that occur at order $p^6$ in a vector meson
dominance model have been obtained in \cite{avref}. They are of the
form
\begin{equation}
{\cal L}_V = {G_8 \alpha^{\phantom{l}}_{EM} \over 4 \pi} {4 a^{\phantom{l}}_V \over M^2_K}
\biggl(\partial^\alpha K_L \partial_\alpha \pi^0
F^{\mu\nu}F_{\mu\nu} + 2 \partial_\alpha K_L \partial^\beta \pi^0
F^{\alpha\mu}F_{\mu\beta} \biggr)
\end{equation}
and, therefore, the prediction of vector meson dominance is that
\begin{eqnarray}
\alpha^{\phantom{l}}_1 &=& -4 a^{\phantom{l}}_V, \nonumber \\
\alpha^{\phantom{l}}_2 &=& 12 a^{\phantom{l}}_V, \nonumber \\
\beta &=& -8 a^{\phantom{l}}_V.
\label{vpatt}
\end{eqnarray}
This prediction is at the heart of Eq.~\ref{cohenan}, and differs from
it only by small additional constants which appear in a particular
regularization scheme for the loop amplitudes \cite{unitco}. Although
this pattern is a firm prediction of vector meson dominance models, a
specific value for $a^{\phantom{l}}_V$ is not. For example, in
Ref.~\cite{avref} the values $a^{\phantom{l}}_V =0.32$ or
$a^{\phantom{l}}_V = -0.32$ can be obtained depending on whether one
uses the so called ``weak deformation model'' or not. This is just
another way of saying that the concept of ``vector meson dominance''
is not uniquely defined for the weak interactions. In addition,
phenomenological treatments of vector mesons such as those of
Ref.~\cite{vmphen} include effects from $\eta -\eta^\prime$ mixing,
which are formally of higher order, but which result in significantly
different ``vector meson'' contributions to $K_L \rightarrow \pi^0
\gamma \gamma$. It is worth mentioning that a quark model estimate of
the parameters $\alpha^{\phantom{l}}_1$, $\alpha^{\phantom{l}}_2$ and
$\beta$ \cite{chiqua} yields the same pattern as in Eq.~\ref{vpatt}
with $a^{\phantom{l}}_V = (N_c /27)g_A^2 (M^2_K/m^2)$ in the notation
of \cite{chiqua}.

In addition to the vector meson exchange contributions, the parameters
$\alpha^{\phantom{l}}_1$, $\alpha^{\phantom{l}}_2$ and $\beta$ may receive contributions from
the exchange of scalar and tensor resonances. The effect of scalar
resonances near 1 GeV has been found to be small \cite{sfscalar}, and
we include it here for completeness. Moreover,
we sidestep the issue of a possible scalar resonance in the
vicinity of 500~MeV because the physics of this broad enhancement
in the $J=I=0$ $\pi \pi$ scattering amplitude is, to a large extent,
already included in the treatment of the pion loops.
We concentrate instead in resonances near 1~GeV
such as the $f_0(980)$, and
take the simplest form for the scalar-pion and scalar-photon
interactions \cite{drvres}
(we use $U$ as in the notation of Gasser and Leutwyler \cite{gl}):
\begin{equation}
{\cal L_S} = g_\pi S {\rm Tr} \biggl(D^\mu U D_\mu U^\dagger\biggr)
+ {\alpha^{\phantom{l}}_{EM} \over 4 \pi} g_\gamma S F^{\mu \nu} F_{\mu \nu}.
\end{equation}
We have not included a coupling of the scalar field proportional
to light quark masses because it does not contribute to
$K_L \rightarrow \pi^0 \gamma \gamma$, and because there is not enough
experimental information on scalar-meson decays to extract it.

The coupling $g_\pi$ can be determined from the decay
width of the scalar into two pions. Adding the charged and
neutral modes we obtain
\begin{equation}
\Gamma(S \rightarrow \pi \pi) = {3 \over 8 \pi f_\pi^4}
\sqrt{1-4 r_{\pi s}^2}g_\pi^2 M^3_S
\biggl(1 - 2r^2_{\pi S} + 4r^4_{\pi S}\biggr),
\end{equation}
with $r_{\pi S} = M_\pi/M_S$.
If we identify the scalar meson with the $f_0(980)$, and use
the particle data book figures $B(f_0 \rightarrow \pi^+\pi^-) = 2/3$,
$B(f_0 \rightarrow \pi^0\pi^0) = 1/3$,
\cite{pdb} and the NOMAD result $\Gamma(f_0) = 35 \pm 12$~MeV
\cite{nomad} we find $g_\pi \sim \pm 5$~MeV (we cannot decide the sign
ambiguity from the experimental rates).

The width for the scalar-meson decay into two photons allows us
to determine $g_\gamma$. We find for the width
\begin{equation}
\Gamma(S \rightarrow \gamma \gamma) = \biggl({\alpha^{\phantom{l}}_{EM}\over 4 \pi}
\biggr)^2{g^2_\gamma M_S^3 \over 4 \pi}.
\end{equation}
If again we identify the scalar with the $f_0(980)$ and use the
particle data book value $\Gamma(f_0 \rightarrow \gamma \gamma)
= 0.39^{+0.10}_{-0.13} \times 10^{-3}$~MeV \cite{pdb}, we find
$g_\gamma \sim \pm 3.9 \times 10^{-3}$~MeV$^{-1}$.

\begin{figure}[!htb]
\begin{center}
\epsfxsize=15cm
\centerline{\epsffile{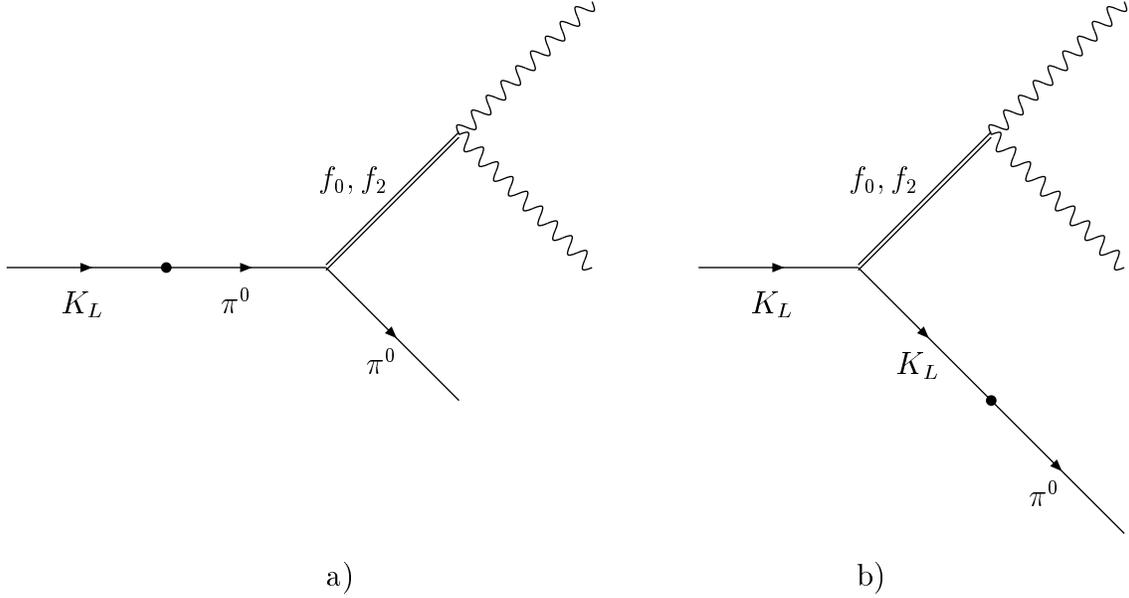}}
\end{center}
\caption{Scalar- and tensor-meson resonance Feynman diagrams
contributing to $K_L \rightarrow \pi^0 \gamma \gamma$. The dots in a) and b)
represent flavor-changing mass-insertions in the incoming and
outcomimg particles, respectively \cite{locpt,rafael,dgh}.}
\label{fig:feyn}
\end{figure}

Collecting these results we finally obtain for the contribution
of the scalar $f_0(980)$ to $K_L \rightarrow \pi^0 \gamma \gamma$ (see
Fig.~\ref{fig:feyn}):
\begin{equation}
\alpha^{\phantom{l}}_1 = -\alpha^{\phantom{l}}_2 = - 16 g_\pi g_\gamma{M^2_K \over M^2_S}
\sim \mp 0.08, \qquad \beta = 0.
\end{equation}

In a similar manner we can determine the contribution from a
tensor meson. A simple look at the low energy data for the
reaction $\gamma \gamma \rightarrow \pi^0 \pi^0$ \cite{cball}
suffices to motivate the potential importance of the $f_2(1270)$
for our amplitudes through diagrams such as those in Fig.~\ref{fig:feyn}.
Following Ref.~\cite{drvres} we write the lowest
order couplings of a tensor meson $T_{\mu\nu}$ to pions and photons as
\begin{equation}
{\cal L}_T= h_\pi T^{\mu\nu}{\rm Tr}\biggl(D_\mu U
D_\nu U^\dagger \biggr) + {\alpha^{\phantom{l}}_{EM}\over 4 \pi}h_\gamma T^{\mu\nu}
F_{\mu\alpha}F_\nu^\alpha.
\end{equation}
For the inclusive width of the tensor meson into two pions, and
following Ref.~\cite{pilkuhn} for the description of the spin 2 states,
we obtain
\begin{equation}
\Gamma(T \rightarrow \pi\pi) = {3 h_\pi^2 M_T^3 \over 240 \pi f_\pi^4}
\biggl(1-4m^2_\pi/M_T^2\biggr)^{5/2}.
\end{equation}
For the decay width of the tensor meson into two photons we find
\begin{equation}
\Gamma(T \rightarrow \gamma \gamma) = \biggl({\alpha^{\phantom{l}}_{EM}\over 4 \pi}\biggr)^2
{h_\gamma^2 M_T^3 \over 80 \pi}.
\end{equation}
Identifying the tensor meson with the $f_2(1270)$ and using the
particle data book values for mass and partial widths \cite{pdb},
we obtain $h_\pi \sim \pm 40$~MeV and $h_\gamma \sim \pm 0.03$~MeV$^{-1}$.

The tensor ($f_2$) contribution to the parameters $\alpha^{\phantom{l}}_1$,
$\alpha^{\phantom{l}}_2$ and $\beta$ can be read from the interaction that
results after the tensor meson has been integrated out
\begin{equation}
{\cal L}_T = {G_8 \alpha^{\phantom{l}}_{EM} \over 4 \pi} {4 h_\pi h_\gamma \over M^2_T}
\biggl( {2 \over 3} \partial^\alpha K_L \partial_\alpha \pi^0
F^{\mu\nu}F_{\mu\nu} + 2 \partial_\alpha K_L \partial^\beta \pi^0
F^{\alpha\mu}F_{\mu\beta} \biggr).
\end{equation}
The resulting contributions are:
\begin{eqnarray}
\alpha^{\phantom{l}}_1 &=& {4 \over 3} h_\pi h_\gamma
{M^2_K\over M^2_T} \sim \pm 0.25, \nonumber \\
\alpha^{\phantom{l}}_2 &=& - {28 \over 3} h_\pi h_\gamma
{M^2_K\over M^2_T} \sim \mp 1.7, \nonumber \\
\beta &=& 8  h_\pi h_\gamma
{M^2_K\over M^2_T} \sim \pm 1.5.
\end{eqnarray}

We summarize our results in Table~\ref{tab}.

\begin{table}[htb]
\centering
\begin{tabular}{|c|c|c|c|c|c|} \hline
& Vector ($a_V$ = $\pm$ 0.32) & Scalar & Tensor & Our Best Fit & Best Fit
$a_V$ \\
\hline
\hline
$\alpha^{\phantom{l}}_1$ & $\mp$ 1.2 & $\mp$ 0.08 & $\pm$ 0.25 & 0 & 3.8 \\
\hline
$\alpha^{\phantom{l}}_2$ & $\pm$ 3.6 & $\pm$ 0.08 & $\mp$ 1.7 & 1.7 & --12 \\
\hline
$\beta$ & $\mp$ 2.4 & 0 & $\pm$ 1.5 & --5 & 7.5 \\
\hline
\end{tabular}
\vskip 0.1 in
\caption{A comparison of parameters for $K_L\rightarrow\pi^0\gamma\gamma$
for various contributions discussed in the text. We contrast these
contributions with our best three-parameter fit, as well as with our
best fit within the VMD ansatz.}
\label{tab}
\end{table}

\section{Conclusion}

We expect new data for $K_L \rightarrow \pi^0 \gamma\gamma$ from KTeV
and NA48 in the near future, and this makes a reanalysis of this mode
timely. We have argued that the new results should not be analyzed in
terms of the vector meson dominance ansatz, but rather in a model
independent way, and that this entails the use of three parameters.
These three parameters are related to the three a priori undetermined
counterterms entering the amplitude, as shown in Eqs. \ref{count},
\ref{cpara}.

To illustrate the previous point we have re-examined the fit to the
1999 KTeV data. We find that the general, three-parameter fit is
slightly better than the old fit in terms of $a_V$, and we show our
results in Fig.~\ref{fig:fit}. The difference between the two
procedures appears to be small in the $K_L \rightarrow \pi^0\gamma
\gamma$ spectrum. Nevertheless, it leads to significantly different
predictions for the CP-conserving component of $K_L \rightarrow \pi^0
e^+ e^-$, which can be seen in Eqs.~\ref{comp}, \ref{comp2}. New data,
with higher statistics, should be able to better distinguish the two
cases.

As a further motivation for abandoning the usual parameterization,
we have also shown that the $f_2(1270)$ tensor meson can yield an
important contribution to the counter-terms, and that this
contribution cannot be cast in the one-parameter framework of
vector meson dominance.

\section*{Acknowledgments}

\noindent This work  was supported in part by DOE under contact number
DE-FG02-01ER41155. We are grateful to Gino Isidori for discussions
on this matter. We also thank Rainer Wanke (NA48), Peter Shawhan
and Yee-Bob Hsiung (KTeV) for comments on the original manuscript.

\section*{Appendix}

To compute an estimate of the errors involved in our fits we calculate
the range of variation of the three parameters from our ``best fit'' $\chi^2_{\rm
min}$ = 37 to $\chi^2_{\rm min}$ + 1 (corresponding to one standard deviation), obtaining:
\begin{eqnarray}
-2.0 <& \alpha^{\phantom{l}}_1& < 1.9, \nonumber \\
0.8 <& \alpha^{\phantom{l}}_2& < 2.5, \nonumber \\
-5.3 <& \beta& < -4.5.
\end{eqnarray}

The ``central'' values give roughly a linear equation:
\begin{equation}
\alpha^{\phantom{l}}_2 + 0.29\, \alpha^{\phantom{l}}_1 = 1.65.
\end{equation}
This is consistent with $z-r_\pi^2\sim 0.3$ dominating any 
$z$ dependence.

\noindent The above values are to be compared with the ones given in Eq.~\ref{3pfit}:
\begin{eqnarray}
\alpha^{\phantom{l}}_1 &=& 0, \nonumber \\
\alpha^{\phantom{l}}_2 &=& 1.7, \nonumber \\
\beta &=& -5.
\end{eqnarray}

\begin{figure}[!htb]
\begin{center}
\epsfxsize=15cm
\centerline{\epsffile{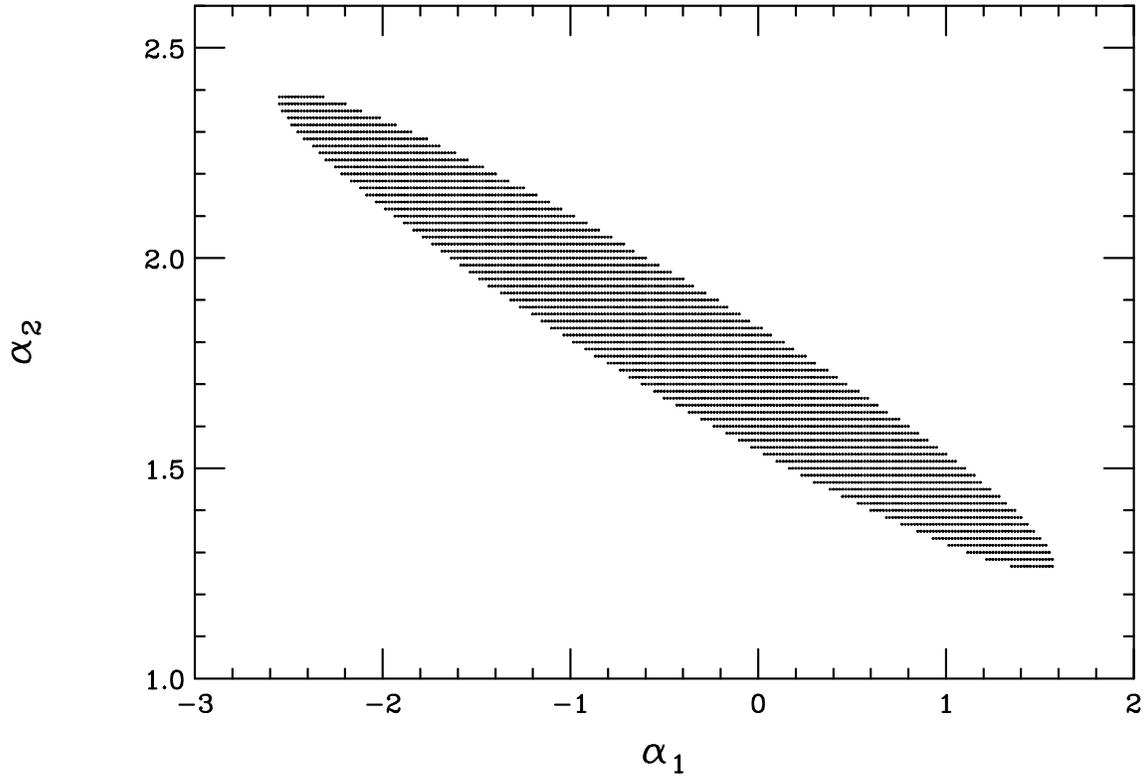}}
\end{center}
\caption{Scatter plot of the parameter space allowed for
$\alpha_1$, $\alpha_2$ with a fixed $\beta=-5$, within one sigma from
our $\chi^2_{\rm min}$.}
\label{fig:fiter}
\end{figure}

\noindent In Fig.~\ref{fig:fiter} we present a one-sigma plot of the
parameter space for $\alpha_1$ and $\alpha_2$ with a fixed $\beta =
-5.0$\,.

It is possible to redo the calculation keeping each time one parameter
variable and the other two fixed at the values of Eq.~\ref{3pfit}. In
this case they are much more constrained:

\begin{eqnarray}
-0.5 < &\alpha^{\phantom{l}}_1& < 0.6, \nonumber \\
 1.5 < &\alpha^{\phantom{l}}_2& < 1.8, \nonumber \\
-5.1 < &\beta& < -4.9.
\end{eqnarray}

There exists another region in the parameter space where $\chi^2$
is within one sigma from $\chi^2_{\rm min}$:

\begin{eqnarray}
-0.6 <& \alpha^{\phantom{l}}_1& < 0.8, \nonumber \\
-12.2 <& \alpha^{\phantom{l}}_2& < -11.8, \nonumber \\
8.4 <& \beta& < 8.7.
\end{eqnarray}
Note that the one-parameter fit in terms of $a^{\phantom{l}}_V$ lies
closer to this second region.

\begin{figure}[!htb]
\begin{center}
\epsfxsize=15cm
\centerline{\epsffile{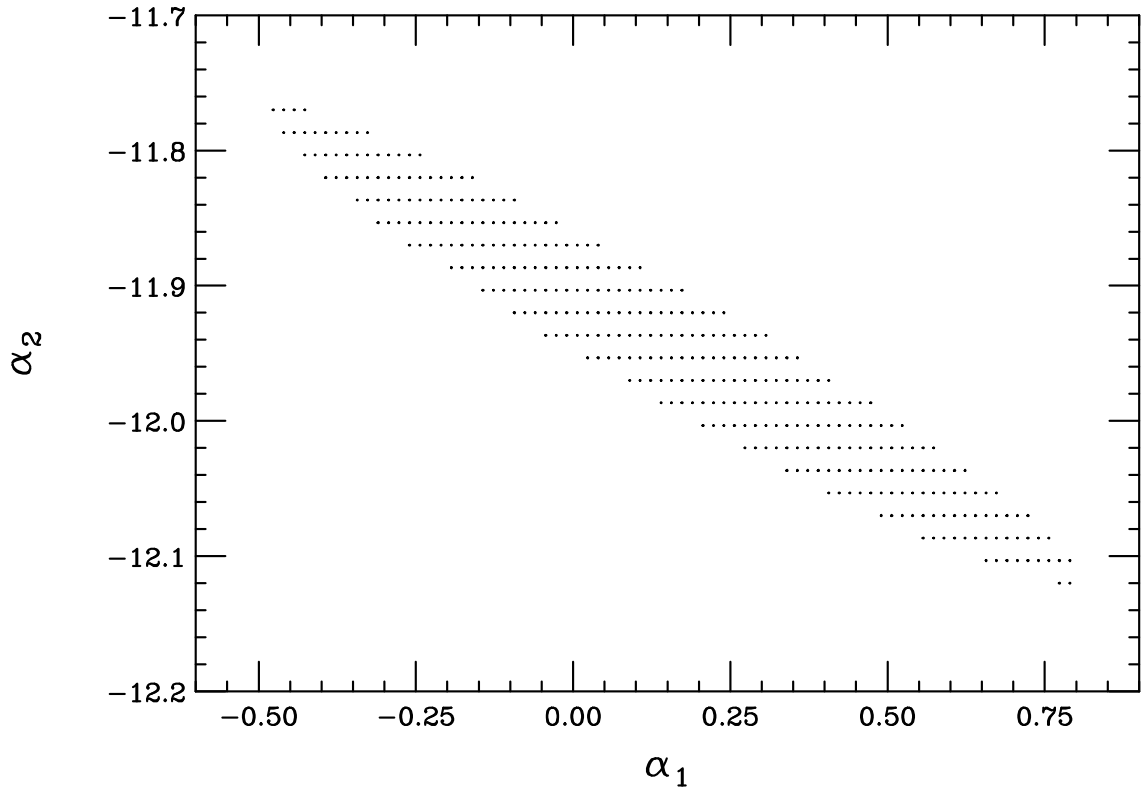}}
\end{center}
\caption{Same as Fig.~\ref{fig:fiter} with $\beta=8.55$.}
\label{fig:fiter2}
\end{figure}

Fig.~\ref{fig:fiter2} is analogous to Fig.~\ref{fig:fiter} assuming $\beta = 8.55$,
and is consistent with the ``central values'' linear equation
\begin{equation}
\alpha^{\phantom{l}}_2 + 0.28 \, \alpha^{\phantom{l}}_1 = -11.9.
\end{equation}

\end{document}